\documentclass[aps,pra,twocolumn,reprint,amssymb,floats,superscriptaddress]{revtex4-2}
\usepackage{graphicx}
\usepackage{bm}
\usepackage{amsmath}
\usepackage{amssymb}
\usepackage{amsfonts}
\usepackage{euscript}
\usepackage{verbatim}
\usepackage{setspace}
\usepackage{xcolor}
\usepackage{amsfonts}
\usepackage{braket}
\usepackage{multirow}
\usepackage[unicode=true]{hyperref}
\usepackage{listings}
\usepackage{adjustbox}

\usepackage{tikz}
\usetikzlibrary{quantikz}

\newcommand{\be}{\begin{equation}}
\newcommand{\ee}{\end{equation}}
\newcommand{\bea}{\begin{eqnarray}}
\newcommand{\eea}{\end{eqnarray}}

\begin{document}
\title{Simulating long-range coherence of atoms and photons in quantum computers}
\author{Emanuele G. Dalla Torre}
\affiliation{Department of Physics and Center for Quantum Entanglement Science and Technology, Bar-Ilan University, 52900 Ramat Gan, Israel}
\affiliation{Superconducting Quantum Materials and Systems Center (SQMS), Fermi National Accelerator Laboratory, Batavia, IL 60510, USA}
\author{Matthew J. Reagor}
\affiliation{Rigetti Computing, 775 Heinz Avenue, Berkeley, CA 94710}
\affiliation{Superconducting Quantum Materials and Systems Center (SQMS), Fermi National Accelerator Laboratory, Batavia, IL 60510, USA}

\begin{abstract}
Lasers and Bose-Einstein condensates (BECs) exhibit macroscopic quantum coherence in seemingly unrelated ways. Lasers possess a well-defined global phase and are characterized by large fluctuations in the number of photons. In BECs of atoms, instead, the number of particles is conserved and the global phase is undefined. Here, we present a unified framework to simulate lasers and BECs states in gate-based quantum computers, by mapping bosonic particles to qubit excitations. Our approach relies on a scalable circuit that measures the total number of particles without destroying long-range coherence. We introduce complementary probes to measure the global and relative phase coherence of a quantum state, and demonstrate their functionality on a Rigetti quantum computer. Our work shows that particle-number conservation enhances long-range phase coherence, highlighting a mechanism used by superfluids and superconductors to gain phase stiffness.
\end{abstract}

\maketitle

\newcommand \mysection[1]{{\it #1 -- }}

\mysection{Introduction}
One of the fundamental principles of quantum mechanics is that the number and the phase operators are canonical conjugates \footnote{See Sec.~\ref{sec:commutation} of the SM for a simple proof.}. Accordingly, in a system with a fixed number of particles, such as a physical gas or liquid, the global phase operator has maximal uncertainty. This simple observation seems to contradict our basic understanding of Bose-Einstein condensates (BECs), superfluids, and superconductors, where phase coherence emerges in spite of particle-number conservation. This apparent contraddition is resolved by noting that, in these systems, the phase coherence is imprinted in {\it relative} degrees of freedom, which are uneffected by this uncertainty principle. As a consequence, to probe the phase coherence of a BEC, it is always necessary to perform an interferometric experiment between two parts of the system \footnote{See f.e.~Refs.~\cite{hagley1999measurement,inguscio1999bose,bloch2000measurement,shin2004atom,schumm2005matter} for more details about interference experiments with BEC of ultracold atoms and Ref.~\cite{marelic2016spatiotemporal,damm2017first} for BEC of photons.}. 

Long range coherence also occurs in systems that do not conserve the total number of particles. The simplest example is a laser, where the total number of photons fluctuates, and the global phase is well defined and can be probed directly. An intermediate situation between BECs and lasers is offered by BECs of light in optical cavities, such as the BEC of exciton-polaritons \cite{kasprzak2006bose,deng2010exciton} and the BEC of photons in dye molecules \cite{klaers2010thermalization,klaers2012statistical,walker2018driven}. The cavities increase the lifetime of the photons, making their number a quasi-conserved quantity and allowing them to reach a BEC state.
 Unlike BECs of atoms, in BECs of lights the total number of particles has large fluctuations due to cavity losses and external reservoirs \cite{kocharovsky2006fluctuations,schmitt2014observation}. The relation between these three many-body states (lasers, BECs of atoms, and BECs of light) has been the subject of a long-standing debate \cite{kirton2013nonequilibrium,carusotto2013quantum,byrnes2014exciton,wang2019theory,kirton2019introduction}, in part due to the absence of a single platform where they can be studied on equal footing. Here, we show how to use gate-based quantum computers to create these states, and study their coherence and fluctuations, see Table \ref{table1}. 


\begin{table}[b]
	\begin{tabular}{|l|l|l|l|}
		\hline
		Many-body state & qubit state  & number fluct. & coherence \\
		\hline
		Laser & $|\rm coherent\rangle$  & $\langle S_z^2\rangle = N/4$ & $\langle S_x\rangle=N/2$\\
		\hline
		BEC of light & $|\rm dephased\rangle$ & $\langle S_z^2\rangle = N/4$ & $C_N^{(2)} \approx 1/4$\\
		\hline
		BEC of atoms & $|\rm projected\rangle$  & $\langle S_z^2\rangle = 0$ & $C_N^{(2)} \approx 1/4$ \\
		\hline
		Thermal & $|\rm noisy\rangle$  & $\langle S_z^2\rangle = N/4$ & $C_N^{(2)} \approx 0$\\
		\hline
	\end{tabular}
	\caption{Many-body states of superconducting circuits, used to simulate  long-range coherence: $|\rm dephased\rangle$ is obtained by measuring the total $S_z$; $|\rm  projected\rangle$ state by post-selecting the $S_z=0$ result; $|\rm noisy\rangle$ is a statistical mixture of $|0\rangle$ and $|1\rangle$.}
	\label{table1}
\end{table}

\mysection{Coherent, dephased, and projected states} Following a common approach~\cite{auerbach2012interacting}, we map the presence (absence) of an particle to  the $|0\rangle$ ($|1\rangle$) state of a qubit. A laser state is mapped to the spin coherent state of $N$ qubits,
\begin{align}
{|{\rm coherent}\rangle } = \prod_{n=1}^N \frac1{\sqrt{2}}\left(|0\rangle_n+e^{i\theta_n}|1\rangle_n\right).
\end{align}
Here, the variable $\theta_n$ is the direction of the $n$th spin in the XY plane and corresponds to the local phase of the coherent state. In the case of $\theta_n=0$ all the spins point in the X direction and ${|{\rm coherent}\rangle }$ is an eigenstate of the spin operator $S_x = \sum_i \sigma^x_i$, with maximal eigenvalue $N/2$. In analogy to a laser, this state has a well-defined global phase, $\theta=0$ and its number of particles $n = S_z - N/2$ has large quantum fluctuations, $\delta n = \delta S_z = \sqrt{N}/2$. 


We now move to BECs of light, which can be prepared in a state with large fluctuations of both the global phase and the total number of particles \cite{kocharovsky2006fluctuations,schmitt2014observation}. This state can be obtained by measuring $S_z$ and keeping all possible outcomes of the measurement. The resulting state is described by 
\begin{align}
\rho_{\rm dephased} = \sum_{s=-N/2}^{N/2} \delta_{S_z,s}|\rm coherent\rangle \langle \rm coherent|,
\end{align}
where $\delta_{S_z,s}$ is a projection operator. For simplicity, we will denote this state by $|\rm dephased\rangle$, in spite of being mixed.  In BEC of light, the total number of particles is measured by the external baths or reservoirs, while in our simulator we will achieve this goal using ancilla qubits. Interestingly, the same state can be obtained by considering $|\rm coherent\rangle$ with an homogeneous $\theta_n=\theta$ and setting the global phase $\theta$ to be a random variable with uniform distribution in $[0,2\pi)$. 
In what follows, we will show that this state is nevertheless characterized by long-range phase coherence.

Finally, to simulate a BEC of atoms, we project the state to a subspace with a well-defined number of particles. For concreteness, we assume that the $N$ is even and consider the projection over the subspace with $N/2$ atoms, or equivalently $S_z=0$,
\begin{align}
	|{\rm projected}\rangle = {\mathcal{A}} \delta_{S_z,0}|{\rm coherent}\rangle,
\end{align}
where $\mathcal{A}$ is a normalization factor. In the case of $N=2$ qubits, one has 
$|{\rm coherent}\rangle =  \frac12\left(|0\rangle+e^{i\theta_1}|1\rangle\right)\left(|0\rangle + e^{i\theta_2}|1\rangle\right)$.
By post-selecting the state with $S_z=0$, one obtains the Bell state
$|{\rm projected}\rangle =  \left(|01\rangle+e^{i(\theta_2-\theta_1)}|10\rangle\right)/\sqrt{2}.$
This state is invariant under the global phase rotation $\theta_i\to\theta_i+\Delta\theta$, but retains the information about the relative phase $\theta_1-\theta_2$. Interestingly, this procedure allows one to create entanglement between two qubits without having them interact directly, as proposed in Ref.~\cite{motzoi2015continuous} and experimentally realized with superconducting circuits in Ref.~\cite{dickel2018chip}.  Here, we aim at extending this analysis to large numbers of particles and studying their long-range coherence. 

\mysection{Probing long-range coherence}
In analogy to the case of a laser, the coherence of $|\rm coherent\rangle$ can be directly measured by probing the expectation values of the spin operator $\langle S_x\rangle = N/2$. In contrast, for the BEC states, the global phase is undefined and $\langle S_x\rangle =0$. We now discuss two complementary methods to probe the phase coherence of a these states. 

The first method targets the phase correlations between the qubits. In the two-qubit Bell state, the relative phase is probed by a finite expectation value of the operator $\langle \sigma_1^+\sigma_2^- + \sigma_2^+\sigma_1^-\rangle = 2\cos(\theta_1-\theta_2)$. The many-body generalization of this operator is the two-point correlator
\begin{align}
C_N^{(2)} = \frac 1{2N^2} \langle S^+ S^- + S^- S^+\rangle,
\label{eq:CN2}
\end{align}
with $S^\pm=\sum_i \sigma^\pm_i$. In this work we consider states with $\langle S_z\rangle=0$, for which $C_N^{(2)}=\langle S_x^2+S_y^2\rangle/N^2$  \footnote{The operator $C_N^{(2)}$ can be easily probed by (i) measuring $\sigma_n^x$ of each qubit, (ii) computing the total spin $S_x$ and squaring the result, (iii) averaging over many realizations (shots) and (iv) repeating the procedure for $S^y$. The correlator (\ref{eq:CN2}) is the same used in quantum magnetism to detect mean-field XY ferromagnetic order.} From a theoretical perspective, a quantum state is phase coherent if $C_N^{(2)}$ remains finite in the limit of $N\to\infty$. In the technical language, this situation corresponds to the spontaneous breaking of the $U(1)$ gauge symmetry associated with particle-number conservation, also known as off-diagonal long-range order \cite{yang1962concept}. Note that $C_N^{(2)}>(\langle S_x\rangle)^2$, and hence having $\langle S_x\rangle$ that scales with $N$ is a sufficient condition for long-range coherence. An example of a state without long-range coherence is offered by the state $|\rm noisy\rangle$ obtained by randomly preparing each qubit in the $|0\rangle$ and $|1\rangle$ states. In this state, $\langle S_x^2\rangle = \langle S_y^2\rangle =N/2$, and $C_N^{(2)}$ tends to 0 at large $N$.

\begin{figure*}
	\centering
	\begin{tabular}{c c c c}
		(a) $|{\rm coherent}\rangle$&(b) $|{\rm dephased}\rangle$ &(c) $|{\rm projected}\rangle$ & (d) $|{\rm noisy}\rangle$\\
		\includegraphics[width=4.5cm]{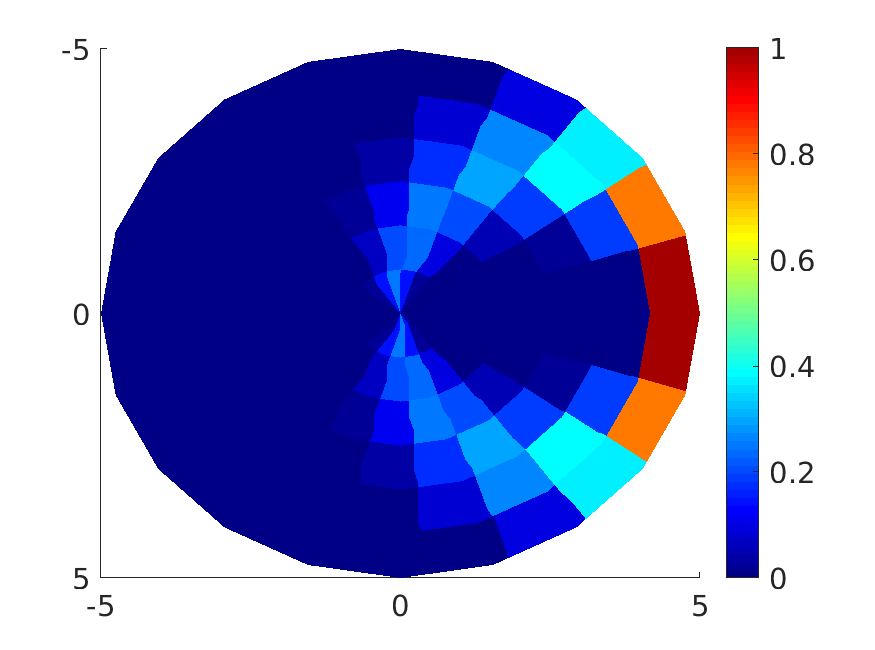}&
		\includegraphics[width=4.5cm]{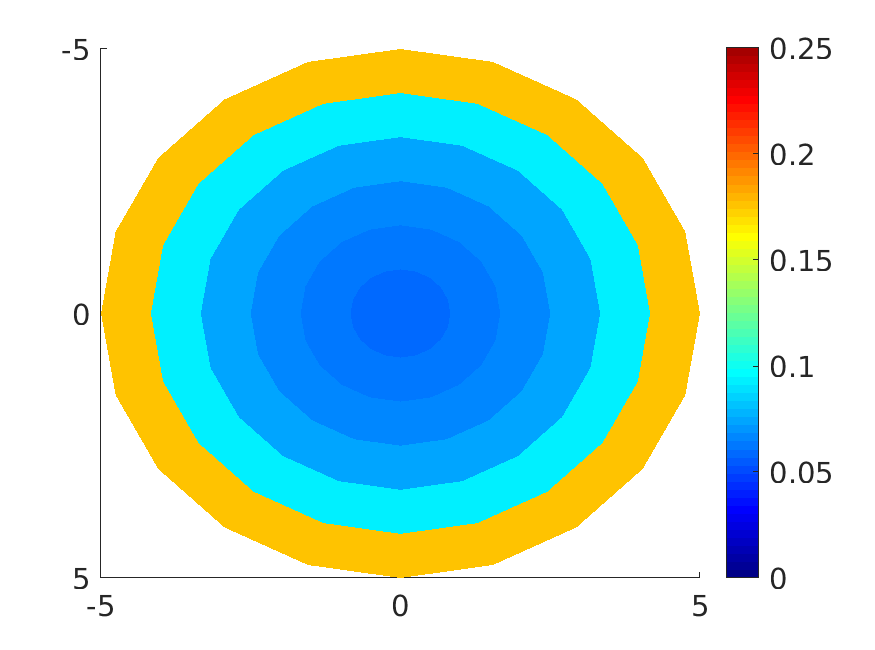}&
		\includegraphics[width=4.5cm]{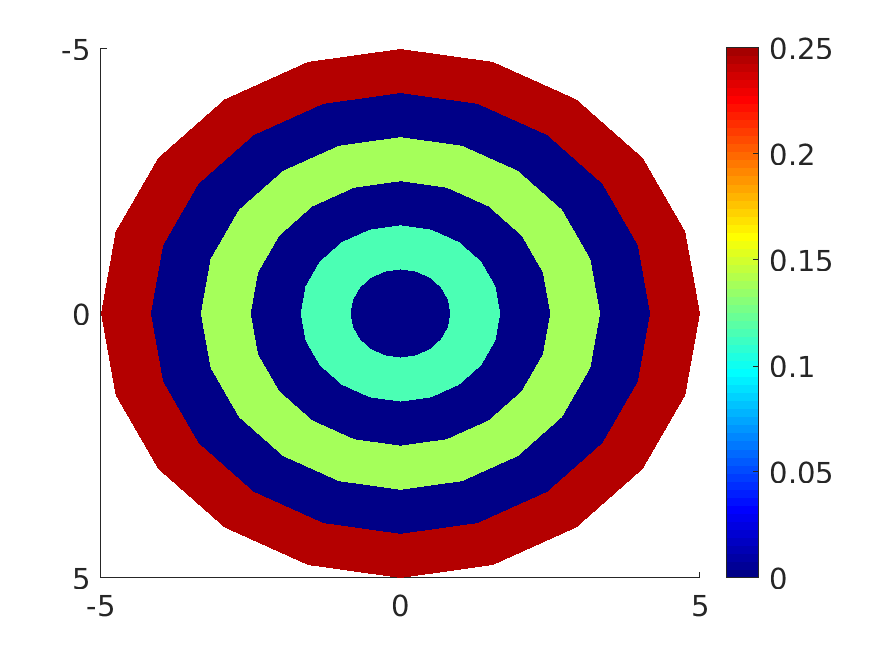}&
		\includegraphics[width=4.5cm]{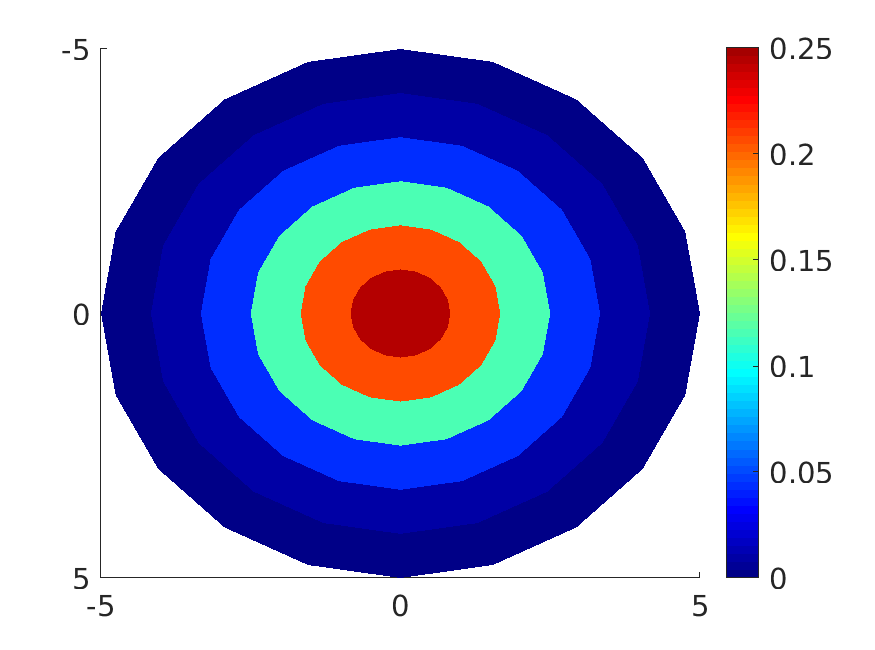}
	\end{tabular}
	\vspace{-0.5cm}
	\caption{Full counting statistic of the operator $S_\theta$, defined in Eq.~\ref{eq:Stheta}, for a system of $N=10$ qubits. The polar direction corresponds to $\theta$ and the radial direction to the size of $S_\theta \in [-N/2,N/2]$. The color coding represents the probability of observing a specific value of $S_\theta$ for a fixed $\theta$. In states with long-range coherence (a-c) the graph is peaked at large absolute values of $S_\theta$.
	}
	\label{fig:coherence}
\end{figure*}

We now show that the three states $|\rm coherent\rangle$, $|\rm dephased\rangle$ and $|\rm projected\rangle$ have long-range coherence. Because these states belong to the fully symmetric subspace with $S^2=N(N+2)/4$, their coherence is related to the fluctuations of $S_z$ through $C_N^{(2)}=\langle S_x^2\rangle +\langle S_y^2\rangle = \langle S^2 \rangle - \langle S_z^2\rangle$. In $|{\rm coherent}\rangle$, the qubits are uncorrelated, such that $\langle S_z^2\rangle=\sum_i \langle (\sigma^z_i)^2\rangle = N/4$ and
\begin{align}
C^{(2)}_{N,{\rm coherent}}= \frac{N+1}{4N},\label{eq:CNcoherent}
\end{align}
The dephasing process that leads to $|\rm dephased\rangle$ preserves $S_z$ and, hence, leaves the coherence unchanged, 
\begin{align}C_{N,\rm dephased}^{(2)}=C^{(2)}_{\rm N,coherent}.\label{eq:CNdephased}\end{align}
Finally, during the creation of $|{\rm projected}\rangle$, $\langle S_z^2\rangle $ is reduced from $N/4$ to zero. Accordingly, $\langle S_x^2+S_y^2\rangle $ is increased by $N/4$, leading to
\begin{align}
C_{N,{\rm projected}}^{(2)}=\frac{N+2}{4N}.\label{eq:CNprojected}
\end{align}
In the thermodynamics limit ($N\to\infty$), Eqs.~(\ref{eq:CNcoherent})-(\ref{eq:CNprojected}) tend to the same value, $1/4$.  This is a signature of the thermodynamic equivalence of the canonical and grand-canonical ensembles.
Interestingly, for any finite $N$ the projected state is {\it more} coherent than the dephased one, $C^{(2)}_{N,\rm projected}>C^{(2)}_{N,\rm dephased}$. We can explain this effect by noting that in $|\rm dephased\rangle$ there is a finite probability to find a state with no particles ($|S_z=-N/2\rangle$), which does not possess any coherence. In contrast, $|{\rm projected}\rangle $ includes only states with $N/2$ particles and its coherence is the maximal attainable in any quantum state.

We now move to a second method to probe the coherence of the BEC states based on the full counting statistic of physical operators \footnote{See also Refs.~\cite{gritsev2006study,kitagawa2010ramsey,kitagawa2011dynamics,smith2013prethermalization} for the use of full counting statics to study the phase coherence of BEC of ultracold atoms}. As explained above, a BEC state is characterized by $\langle S_x\rangle = \langle S_y\rangle  =0$ along with large values $\langle S_x^2+S_y^2\rangle$. This is possible only if $S_x$ and $S_y$ have bimodal distributions with a high probability of finding large absolute values.  To address the gauge invariance of the state, we probe the spin operator in the $\theta$ direction \footnote{The full counting statistic of $S_\theta$ can be experimentally probed in an efficient way by (i) applying a rotation of $\theta$ around the axis $S_z$, (ii) performing a $\pi/2$ rotation, (iii) measuring $\sigma^z_i=s_i$ of each qubit and (iv) computing their sum $S_\theta=\sum s_i$. See also Sec.~\ref{sec:wigner} of the SM for an alternative method based on Wigner distribution, which however requires a full tomography of the quantum state.}, defined as
\begin{align}
	S_\theta = \cos(\theta)S_x +\sin(\theta)S_y\label{eq:Stheta}.
\end{align}
The full counting statistics of this operator for the four states of Table \ref{table1} in a system with $N=10$ particles is shown in Fig.~\ref{fig:coherence}. The polar coordinate of these graphs corresponds to the global phase of the condensate, and the radial one to the possible values of $S_\theta=-N/2,...,N/2$. The latter is related to the phase coherence of the state by $C_N^{(2)}=\pi^{-1}\int_0^{2\pi} d\theta \langle S_\theta^2\rangle/N^2$. These plots shows that only the state $|\rm coherent\rangle$ has a well defined global phase. The plots of the two BEC states, $|\rm dephased\rangle$ and $|\rm projected\rangle$, are rotational symmetric and do not have a well defined global phase. Their long-range coherence is signaled by the large probability to measure $|S_\theta|=N/2$ (i.e., the outermost rings). In contrast, in the state $|{\rm noisy}\rangle$, the result with the largest probability is $S_\theta=0$, signaling the absence of long-range coherence.

A closer inspection of Fig.~\ref{fig:coherence} reveals that the states $|{\rm dephased}\rangle$ and $|{\rm projected}\rangle$ differ in the radial dependence: the former state is a monotonously increasing function of $|S_\theta|$, while the latter is oscillatory and exactly vanishes for all even outcomes. This selection rule can be used to distinguish between the state $|{\rm dephased}\rangle$ and $|{\rm projected}\rangle$ and has a close analogy to the second-order coherence observed in lasers \cite{arecchi1965measurement} and BECs of atoms \cite{schellekens2005hanbury,ottl2005correlations,hodgman2011direct,dall2011observation,perrin2012hanbury}. In these systems, the zero temperature limit of $g^{(2)}(\tau=0)$ tends to 1, indicating the absence of the Hong-Ho-Mandel (HHM) effect. In contrast, for BECs of light, the zero-temperature limit gives $g^{(2)}(\tau)>1$ \cite{horikiri2010higher,schmitt2014observation}. This difference can be understood by noting that in a BEC of atoms, the particles emanate from the same quantum state, while in a BEC of light the particles fluctuate in and out of the system and $g^{(2)}(\tau=0)$ can probe states with a different history (and is, hence, affected the by HHM effect). A similar argument applies here: in $|{\rm projected}\rangle$ all the particles originate from the same $S_z$ state and a selection rule exists such that the probabilities to observe even values of $S_\theta$ exactly vanish. In contrast, in $|{\rm dephased}\rangle$, the particles originate from state with different $S_z$ and the selection rule does not apply.

\mysection{Algorithms for an ideal quantum computer} The BEC states of atoms and photons can be prepared deterministically by, first, rotating each qubit in the $|+\rangle$ state, giving rise to $|\rm coherent\rangle$ and, then, measuring $S_z$. To obtain the state $|{\rm projected}\rangle$ it is further necessary to post-select to outcomes with $S_z=0$.  Importantly, the probability to find this state is lower bounded by $1/N$, showing that the number of measurements required to obtain a fixed precision grows linearly with the number of qubits (and not linearly with the size of the Hilbert space).

\begin{figure}
	\centering
	\includegraphics[width=0.5\textwidth]{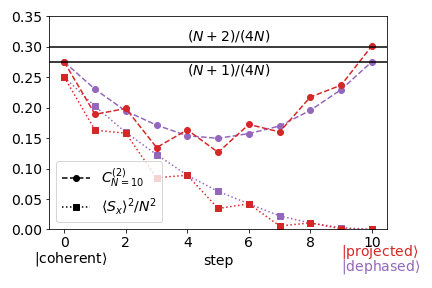}
	\vspace{-0.5cm}
	\caption{Coherence parameters, $C_N^{(2)}$ and $\langle S_x\rangle$, as a function of the number of qubits coupled to the ancillas, for a state with $N=10$ qubits and $N_a=3$ ancillas. The red (purple) curves refer to situation where all the ancilla are (not) postselected, leading to the creation of the state $|\rm projected\rangle$ ($|\rm dephased\rangle$). The horizontal lines correspond to the $C_N^{(2)}$ values for these states, Eqs. (\ref{eq:CNcoherent})-(\ref{eq:CNprojected}).
	}
	\label{fig:coherence_theory}
\end{figure}

The measurement of $S_z$ can be achieved using $N_a={\rm floor}[\log_2(N)]$ ancilla qubits, corresponding to the binary representation of $|S_z|$ (see Ref.~\cite{botelho2022error} for a related algorithm). Each ancilla qubit interacts sequentially with all the qubits, counting the total number of particles. Specifically, we propose to rotate the $a$th ancilla by $\phi_a = 2\pi/2^a$ radians if the qubit is excited, i.e. if a particle is present. At the end of the protocol, the ancilla is rotated by $\phi_a S_z$. The state with $S_z=0$ is obtained by post-selecting the outcomes where all the ancillas are measured in the $|0\rangle$ state. See, for example, Table \ref{table} in \ref{sec:rules} for the case of $N=10$ qubits and $N_a=3$ ancillas.

In Fig.~\ref{fig:coherence_theory} we plot the value of the coherence parameters $C^{(2)}_N$ and $\langle S_x\rangle$ along the process of creating the states $|\rm dephased\rangle$ and $|\rm projected\rangle$, starting from $|\rm coherent\rangle$. At each step, we couple one additional qubit to the ancillas. We observe that $\langle S_x\rangle$ decreases montonously and tends to zero, while $C^{(2)}_N$ has a non-monotonous behavior and tends to Eq.~(\ref{eq:CNprojected}) or Eq.~(\ref{eq:CNdephased}), depending on whether the value of the ancillas are postselected or not. Accordingly, the full counting statistic of the operator $S_\theta$ corresponds to Fig.~\ref{fig:coherence}(b) and (c), respectively. These plots can be used to check that our protocol has successfully prepared the states $|\rm dephased\rangle$ and $|\rm projected\rangle$, corresponding to a BEC of light and a BEC of atoms.

\mysection{Experimental realization}
To realize this protocol in a superconducting quantum computer, we need to overcome several difficulties. First, in these experimental systems each qubit is coupled to at most three other qubits. Hence, a single ancilla cannot be coupled with all the other qubits at the same time. This problem can be solved in a scalable way by considering a linear chain of qubits, such that the ancillas are initially located at one side of the chain. At each step, the ancillas interact with the neighboring qubit and are then swapped with the qubit. As we will see below, these two operations (ancilla rotation and swap) can be combined efficiently.

\begin{figure}
	\includegraphics[width=0.5\textwidth]{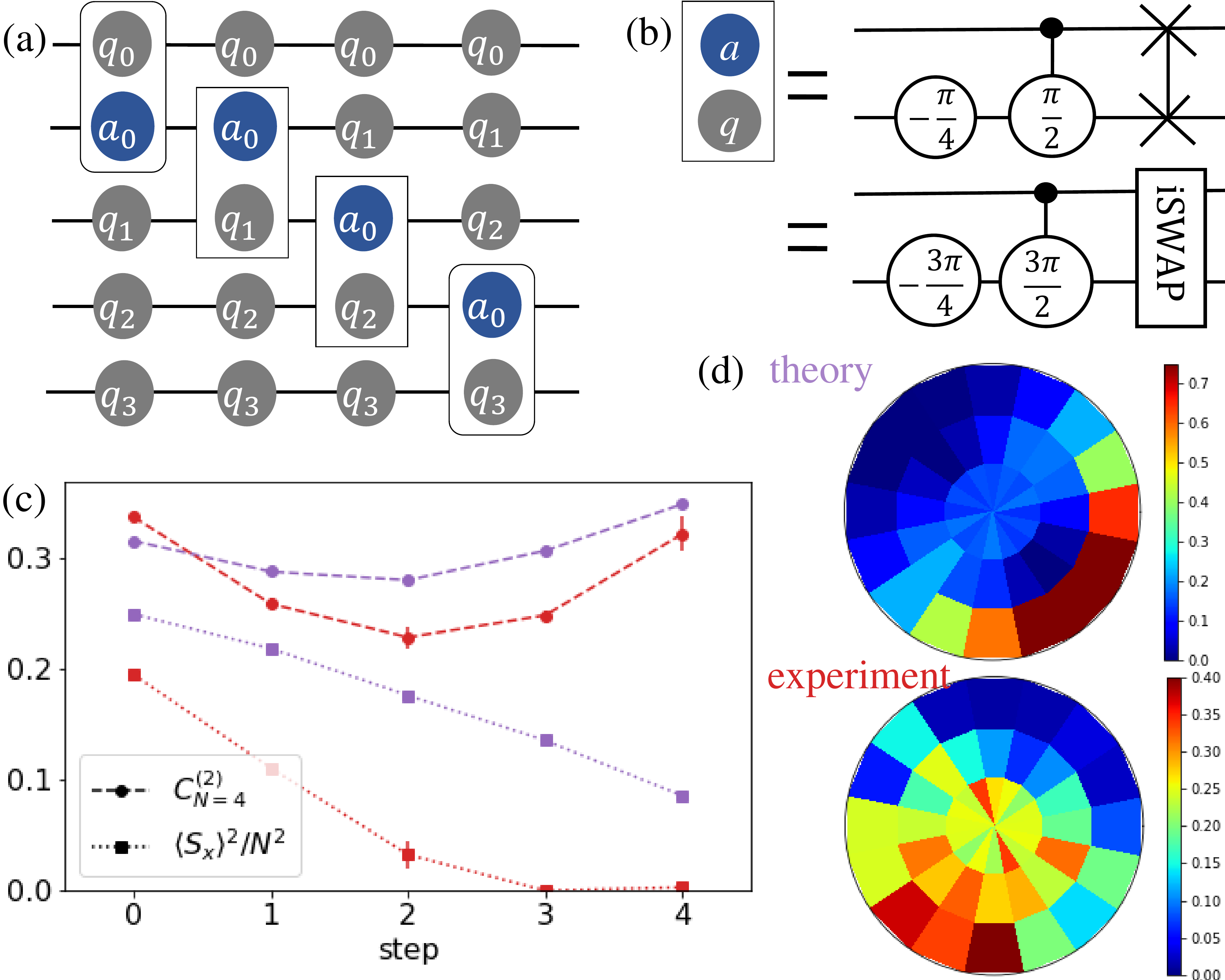}	
	\caption{(a) Protocol required to prepare an approximate BEC state with $N=4$ qubits and $N_a=1$ ancilla. (b) Circuit equivalence used to compile the rotation of an ancilla and swap gates into the Rigetti native gates CPHASE and iSWAP. The circles represent RZ gates.  Two-qubit gates with rounded corners are the same as (b) without SWAP gates. (c) Coherence parameters as a function of the entangling gates: theory (purple) and experiment, scaled by 1.5 (red) (d) Full counting statistics of the final state in theory (upper plot) and experiment (lower plot).}
	\label{fig:experiment}
\end{figure}

The second challenge is to compile the algorithm using as few native two-qubit gates as possible. First, we note that the controlled-rotation CRX($\phi$) gate is usually not a native gate. To overcome this difficulty, we first rotate the ancillas in the XY plane (using, for example a native Hadamard gate), use CPHASE($\phi$) gates to control the ancillas (which are native instructions for Rigetti superconducting processors) and measure the ancillas in the X basis. As mentioned above, the ancilla rotation is often followed by a SWAP operation. However, the product of CPHASE($\phi$) and SWAP is equivalent to the product of two gates, namely CPHASE($\pi+\phi$) and iSWAP (also native for Rigetti), up to single-qubit local gates. See Fig.~\ref{fig:experiment}(b) for the specific case of $\phi=\pi/2$. In total, if we avoid the first and last swap, we obtain a full protocol with $2(N N_a -1)$ two-qubit native gates only. 

For the experimental realization, we consider a simplified version of this protocol, where we use only one ancilla, with $\phi=\pi/2$, and obtain an algorithms with 6 native two-qubit gates, see Fig.~\ref{fig:experiment}(a) \footnote{The ancilla with $\phi=\pi/2$ projects out the state with $S_z=\pm 2$ and, accordingly enhances the coherence of the qubits. See Sec.~\ref{sec:circuit} of the SM for the circuit used to measure $S_\theta$.} The resulting projected state is analogous to $|\rm projected\rangle$ and, in particular, is characterized by a suppressed value of $\langle S_x\rangle \ll N/2$ ($\langle S_x\rangle=1.17$) and an enhanced $C^{(2)}_{4}>C^{(2)}_{4,\rm coherent}$ ($C_4^{(2)}=0.35$). The full counting statistics of $S_\theta$ is plotted in the upper panel of Fig.~\ref{fig:experiment}(d) and shows that global phase spreads over approximately 90 degrees. This is the result of using a single ancilla, instead of the two required to obtain a BEC state with an undefined global phase.

Typical experimental results are shown in Fig.~\ref{fig:experiment}(c) and (d) \footnote{See Sec.~\ref{sec:exp} of the SM for more experimental results.}. In the subplot (c), we have multiplied the experimental results by $50\%$ to account for the lost weight due to gate infidelities and readout errors \footnote{See also Sec.~\ref{sec:mit} of the SM for the implementation of a simple error mitigation scheme \cite{bravyi2021mitigating}.}. The errorbars refer to an average over 3 separate experiments with $N=1000$ shots each. We find that $\langle S_x\rangle$ is a monotonously decreasing function, while $C_N^{(2)}$ is U-shaped, in agreement with the theoretical predictions. The full counting statistics of $S_\theta$ shows that the global phase spreads between $-\pi$ and $-\pi/2$, in contrast to the theoretical one, which spreads between $-\pi/2$ and $0$. We attribute this discrepancy to a coherent dispersive interactions between the qubits, which leads to an effective negative detuning of the qubits and, hence, to a clock-wise rotation of the global phase, which is also observed in the free evolution of the qubits.

\mysection{Conclusion} In this article, we proposed and realized quantum circuits that simulate many-body states with long-range coherence. A key aspect of our work is an efficient algorithm that measures the total number of particles without destroying the phase coherence. Our protocol scales favorably with the number of qubits (the numbers of measurements and gates scales, respectively with $N$ and $N\log_2N$) and can be realized in state-of-the-art quantum computers. Our work clarifies the difference between coherent states and BEC states and shows how to identify them using physical observables. We hope that our work will contribute to the debate on the nature of the BEC of light and its relation to lasing and superradiance.

As a key result, we found that BECs of atoms have larger coherence than coherent states: by reducing the fluctuations in the total number of particles, one obtains a state with a larger phase coherence. An analogous effect occurs in superfluids and superconductors, where local interactions are required to achieve phase stiffness \footnote{See Sec. \ref{sec:stiffness} of the SM for more details about the relation between phase stiffness and suppressed number fluctuations}. To study this connection in a superconducting quantum computer, we plan to prepare a coherent state of many qubits and then couple some ancilla qubits to their local neighbors. By measuring the ancillas and post-selecting the states with a specific outcome, we will obtain an entangled state with long-range phase coherence. These correlations will be immune to local changes of the chemical potential and, hence, potentially provide a noise-free resource for quantum algorithms.

\begin{acknowledgments}
	This material is based upon work supported by the U.S. Department of Energy, Office of Science, National Quantum Information Science Research Centers, Superconducting Quantum Materials and Systems Center	(SQMS) under the contract No. DE-AC02-07CH11359 and by Rigetti Computing. EGDT was supported in part by the Israeli Science Foundation Grant No. 151/19 and 154/19. We acknowledge useful discussions with Jonathan Ruhman, Eleanor Rieffel, Davide Venturelli, Sohaib Alam, Gabriel Perdue, Thomas Iadecola, Alex Hill, Nicolas Didier, Maxime Dupont, Bram Evert, Mark Hodson, and the SQMS Algorithms group. 
\end{acknowledgments}

\bibliography{reference}
\newpage

\onecolumngrid
\newpage

\begin{center}
\begin{large}
Supplemental Material
\end{large}
\end{center}

\setstretch{1.5}

\section{Number and phase operators}
\label{sec:commutation}
In this section, we provide a simple proof of the canonical conjugate of the number and phase operators. This proof is the topic of undergraduate textbooks and is brought here for the sake of completeness. 

Consider the canonically conjugate creation and anihilation operators, $a$ and $a^\dagger$, whose commutation relation is $[a,a^\dagger]=1$. We now define the number operator $n=a^\dagger a$ and the phase operator $\theta$, through $a = e^{i\theta}$. This definition is mathematically sound, although because $\theta$ is not Hermitian, in general it does not correspond to a physical observable. The key step of this demonstration is that
\begin{align}
[a,a^\dagger a]=a ~~~\Rightarrow ~~~[e^{i\theta},n]= e^{i\theta}
\end{align}
The latter identity is satisfied when $[n,\theta]=i$, as one can readily show by performing a Fourier expansion of $e^{i\theta}=\sum_p \theta^p/p!$ and observing that  $[(i\theta)^p,n]=-[n,(i\theta)^p]=p(i\theta)^{p-1}$.

\section{Explicit calculations for two qubits}
\label{sec:2qubits}
In this section, we directly compute the coherence properties of two qubits. These calculations are straightforward and are brought here for completeness only. To highlight the roles of the relative and global phase coherence, we consider the state
\begin{align}
	{|{\rm coherent}_2\rangle } = \prod_{i=1,2} \frac1{\sqrt{2}}\left(|0\rangle_i+|1\rangle_i\right),
\end{align}
In the state $|{\rm coherent}\rangle$ one has $\langle \sigma^x_i\rangle = \cos(\theta_i)$, $\langle \sigma^y_i\rangle = \sin(\theta_i)$. Because the qubits are uncorrelated and $\langle (\sigma^x_i)^2\rangle =1/4$, one obtains $\langle\sigma^x_i\sigma^x_j\rangle = 1/4 \cos(\theta_i)\cos(\theta_j)$. This leads to the spin fluctuations $\langle S_x^2\rangle = 1/2+\cos(\theta_i)\cos(\theta_j)/2$ and $\langle S_y^2\rangle = 1/2+\sin(\theta_i)\sin(\theta_j)/2$, such that 
\begin{align}
	\langle S_x^2 + S_y^2\rangle_{\rm coherent} = 1+ \frac12\cos(\theta_1-\theta_2).
\end{align}

In $|{\rm projected}\rangle$, one has $\langle \sigma_1^+ \sigma_2^-\rangle = e^{i(\theta_1-\theta_2)}$, while $\langle \sigma_1^+ \sigma_2^+\rangle=0$. Hence, $\langle \sigma^x_1\sigma^x_2\rangle = \langle \sigma^y_1\sigma^y_2\rangle = 1/2\cos(\theta_1-\theta_2)$. The spin fluctations are 
\begin{align}
	\langle S_x^2\rangle_{\rm projected}=\langle S_y^2\rangle_{\rm projected} = \frac12+\frac12\cos(\theta_1-\theta_2).\label{eq:appsx2p}
\end{align} 
Using $\langle (S^z)^2\rangle=0$, we find that $\langle S^2\rangle = 1+\cos(\theta_1-\theta_2)$. This result has a simple physical interpretation: in the triplet state ($\theta_1=\theta_2$), the total spin is $s=1$ and $\langle S^2\rangle = s(s+1)=2$. In contrast, in the singlet state $\theta_1-\theta_2=\pi$, $s=0$ and $\langle S^2\rangle = 0$. 

We finally turn to $|{\rm dephased}_2\rangle$. Expectation values in this state can be obtained by averaging over all possible outcomes of the $S_z$ measurement. The case $s_z=0$ corresponds to $|{\rm projected}$, while the state $s_z=\pm1$ correspond to fully polarized states with $S_x^2 = S_y^2 = 1/2$. Hence,
\begin{align}
	\langle S_x^2\rangle_{\rm dephased} = \langle S_y^2\rangle_{\rm dephased} = \frac12+\frac14\cos(\theta_1-\theta_2)\label{eq:appsx2d}.
\end{align}
We find that $\langle S_x^2+S_y^2\rangle$ has the same expectation value as in the $|{\rm coherent}\rangle$. This is not a surprise: measuring $S_z$ does not change the fluctuations of this operator and conserves $S^2$. Hence, $S_x^2+S_y^2$ is unchanged. In the case of $\theta_1=\theta_2=0$, Eqs. \ref{eq:appsx2d} and \ref{eq:appsx2p} coincide with the results of the main text, $C^{(2)}_{2,\rm projected}=1/2$ and $C^{(2)}_{2,\rm dephased}=3/8$.

\section{Wigner function approach}
\label{sec:wigner}
\begin{figure*}[h]
	\centering
	\begin{tabular}{c c c}
		(a) $|{\rm coherent}\rangle$&(b) $|{\rm dephased}\rangle$ &(c) $|{\rm projected}\rangle$\\
		\includegraphics[width=6cm]{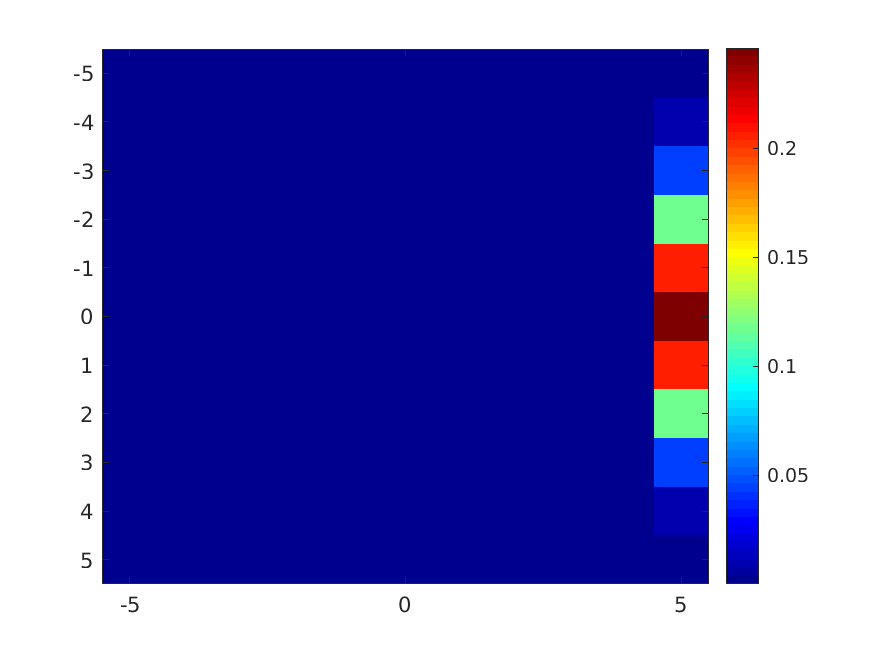}&
		\includegraphics[width=6cm]{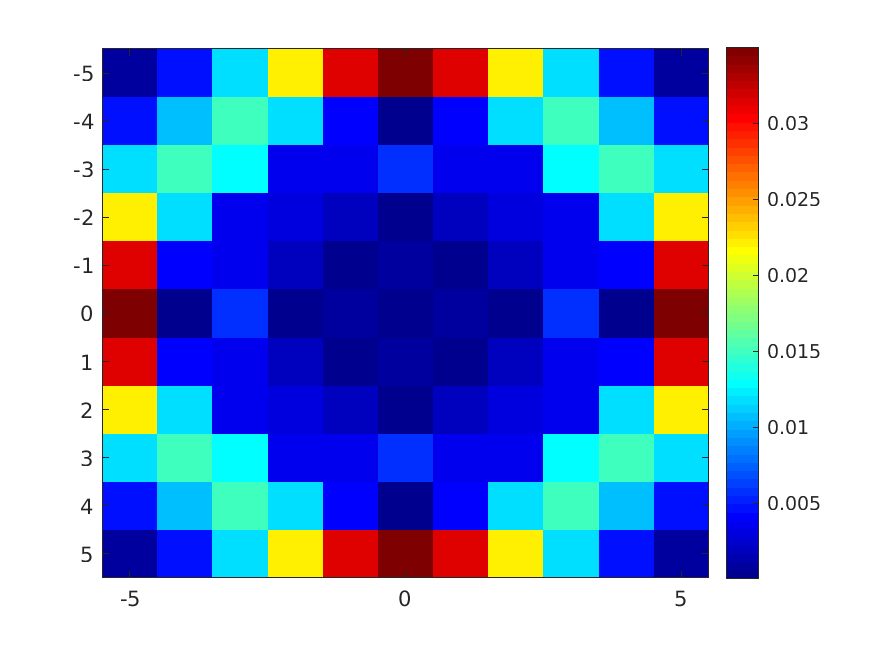}&
		\includegraphics[width=6cm]{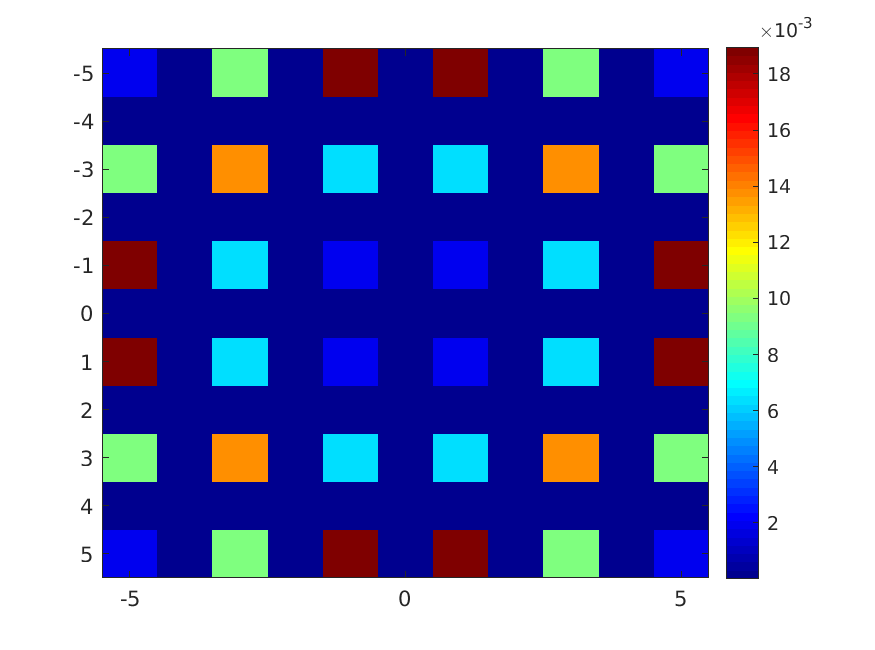}
	\end{tabular}
	\caption{(Wigner functions $W(S_x,S_y)$ defined in Eq.~\ref{eq:wigner} for $N=10$ qubits for the same states as Fig.~\ref{fig:coherence} 
	}
	\label{fig:quick}
\end{figure*}

In this section we present an alternative way to describe the coherence of the states. This probe assumes that one possess the full wave-function of the state, which can be obtained by full state tomography with $3^N$ measurements. We, then, define the Wigner function as
\begin{align}
	W(s_x,s_z)=|\langle \psi|\delta(S_x-s_x) \delta(S_y-s_y)|\psi \rangle|\label{eq:wigner}
\end{align} 
where $\delta(x)=e^{-x^2/\sigma^2}$ and $\sigma\to0$, see Fig.~\ref{fig:quick}. As expected, in the coherent state, the Wigner function is centered around $(S_x,S_y)=(N/2,0)$. In the BEC states, the global phase of the condensate is lost. Nevertheless, most of the weight of the Wigner function is found in areas with large $S_x$ or $S_y$, indicating that, with a high probability, the absolute value of $S^+ = S_x+iS_y$, i.e. $S_x^2+S_y^2$ is macroscopically large. In $|\rm projected\rangle$ the probability to observe even values of $S_x$ and $S_y$ vanishes, in analogy to the selection rule presented in the text.

\section{Using ancillas to probe the number of particles -- a concrete example}
\label{sec:rules}
In the main text we introduced a technique to prepare quantum states with a given particle number using ancillas. The technique involves three mains steps: (i) preparing a spin-coherent state by applying a $\pi/2$ pulse on each qubit individually; (ii) applying the ancillas to the qubits and measuring their state; (iii) post-selecting the state where all the ancillas are in the $0$ state. In this appendix, we provide an example of the function of the ancilla and explain why a logarithmic number of ancillas is sufficient to select the right state.

According to our protocol, the ancilla $a_n$ is rotated by an angle of $S_z\pi/2^n$ radians. As exemplified in Table~\ref{table} for $N=10$ and $N_a=3$, all the states with $S_z\neq 0$ lead to a rotation of $\pi$ of (at least) one ancilla. In this state, the measurement of the ancilla leads deterministically to 1. This result implies that if we select the states where all the ancilla measurement give 0, we project the state to the $S_z=0$ subspace. This protocol can be generalized straightforwardly to larger $N$.

\begin{table}[h]
	\begin{tabular}{|c| c| c| c|}
		\hline
		$S_z$ & $a_0$  & $a_1$ & $a_2$ \\
		\hline
		$0$ & 0 & 0 & 0 \\
		\hline
		$\pm 1$ & \color{red}$\pm \pi$\color{black} & $\pm \pi/2$ & $\pm \pi/4$ \\
		\hline
		$\pm 2$ & $\pm 2 \pi$ & \color{red}$\pm \pi$\color{black} & $\pm 2\pi/4$ \\
		\hline
		$\pm 3$ & \color{red}$\pm \pi$\color{black} & $\pm \pi/2$ & $3\pi/4$  \\
		\hline
		$\pm 4$ & $\pm 4 \pi$ & $\pm 2 \pi$ & \color{red}$\pm \pi$\color{black}\\
		\hline
		$\pm 5$ &  \color{red}$\pm 5\pi$\color{black}  & $\pm 5\pi/2$ & $5\pi/4$\\
		\hline
		$\pm 6$ & $\pm 6\pi$  &  \color{red}$\pm 3\pi$\color{black} & $3\pi/2$\\
		\hline
	\end{tabular}
	\caption{Phase acquired by the ancillas, $\delta \phi_n=S_z\pi/2^n$, for $N=10$ qubits and $N_a=3$ ancillas. When the phase is $(2m+1)\pi$ (red entries), the measurement of the ancilla will deterministically provide the value of 1 and, hence, is projected out by our protocol. The table shows that the projection algorithm leads to the deterministic preparation of the $S_z=0$ state.}
	\label{table}
\end{table}

\section{Quantum circuit used to measure $S_\theta$}
\label{sec:circuit}
Here, we describe the quantum circuit used to measure $S_\theta$ in the case of $N_q=4$ qubits and $N_a=1$ ancilla using $CPHASE$ and $iSWAP=XY(\pi)$ two-qubit gates only. The first ($q_0$) and last ($q_4$) qubits are coupled to the ancilla ($a_1$) through a pair of $R_z(-\pi,4)$ and $CZ(-\pi/2)$ gates, which implement a controlled rotation of $\pi/2$. The other two qubits ($q_1$ and $q_2$) are coupled to the ancillas through the circuit shown in Fig.~\ref{fig:experiment}(b), which implements a controlled rotation, followed by a SWAP gate. The circuit is also draw in Fig.~\ref{fig:circuit}

\begin{lstlisting}
PRAGMA INITIAL_REWIRING "NAIVE"
DECLARE ro BIT[5]
RX(pi/2) 0
RX(pi/2) 2
RX(pi/2) 3
RX(pi/2) 4
RX(pi/2) 1
RZ(-pi/4) 1
CPHASE(pi/2) 0 1
RZ(-3*pi/4) 1
CPHASE(3*pi/2) 2 1
XY(pi) 2 1
RZ(-3*pi/4) 2
CPHASE(3*pi/2) 3 2
XY(pi) 3 2
RZ(-pi/4) 3
CPHASE(pi/2) 4 3
RZ(theta) 0
RZ(theta) 1
RZ(theta) 2
RZ(theta) 4
RX(-pi/2) 0
RX(-pi/2) 1
RX(-pi/2) 2
RX(-pi/2) 4
RX(-pi/2) 3
MEASURE 0 ro[0]
MEASURE 1 ro[1]
MEASURE 2 ro[2]
MEASURE 4 ro[3]
MEASURE 3 ro[4]
\end{lstlisting}

\begin{figure}[h]
	\adjustbox{width=\textwidth,center}{
		\begin{tikzcd}
		q_0 & \gate{R_x(\pi/2)} & \qw & \ctrl{1}  & \qw & \qw & \qw & \qw & \qw & \qw & \qw & \qw   & \gate{R_z(\theta)} & \gate{R_x(-\pi/2)}& \meter{} & \qw & q_0 \\
		a_1 & \gate{R_x(\pi/2)} & \gate{R_z(-\pi/4)} & \gate{R_z(-\pi/2)}  & \gate{R_z(-3\pi/4)} & \gate{R_z(-3\pi/2)}  & \gate[wires=2]{XY(\pi)}  & \qw & \qw & \qw & \qw & \qw  &  \gate{R_z(\theta)} & \gate{R_x(-\pi/2)}& \meter{} & \qw & q_1 \\
		q_1 & \gate{R_x(\pi/2)} & \qw & \qw & \qw & \ctrl{-1} &  & \gate{R_z(-3\pi/4)} & \gate{R_z(-3\pi/2)}  & \gate[wires=2]{XY(\pi)}  & \qw & \qw & \gate{R_z(\theta)} & \gate{R_x(-\pi/2)} & \meter{} & \qw & q_2 \\
		q_2 & \gate{R_x(\pi/2)} & \qw & \qw & \qw & \qw & \qw & \qw & \ctrl{-1} &  & \gate{R_z(-\pi/4)} & \gate{R_z(-\pi/2)} & \gate{R_x(-\pi/2)} & \qw & \meter{} & \qw & a_1 \\
		q_3 & \gate{R_x(\pi/2)} & \qw & \qw & \qw & \qw & \qw & \qw & \qw & \qw & \qw & \ctrl{-1} & \gate{R_z(\theta)} & \gate{R_x(-\pi/2)} & \meter{} & \qw & q_3
	\end{tikzcd}
}

\label{fig:circuit}
\caption{Quantum circuit used to used to measure $S_\theta$ in the case of $N_q=4$ qubits and $N_a=1$ ancilla. The measured value of $S_\theta$ is the sum of $q_0$ to $q_3$ and the projected state is obtained by post-selecting the outcomes with $a_0=0$. }
\end{figure}

\section{More experimental results}
\label{sec:exp}
In table \ref{table:exp}, we report all the experimental results obtained on Aspen-11 and Aspen-M-1. The column "Auto" defines whether the automatic circuit recompilation of pyquil (pyquil\_to\_native\_pyquil) was used. We find that the results obtained using the automatic recompilation were higher (i.e., closer to the theoretical values) than those without recompilation. The last column (``code'') is a unique identifier, which indicates the date of the run: (e) 01/31/2022 (f) 02/03/2022 (g) 03/03/2022 (h) 04/01/2022. Overall, the results of all experiments are comparable to each other. See, for example, Fig. \ref{fig:experiment}, where we generate the same plot as Fig.~\ref{fig:experiment}(c) using a different set of qubits. 

\begin{table}
	\begin{tabular}{|l |l |l| l |l| l|l|}
	\hline
	Device& [$q_0,q_1,q_2,q_3$]& [$a_0$] & Auto & $C^{(2)}_{N}$ &  $\langle S_x\rangle^2/N^2$ & Code \\
	\hline

Aspen-11&[3, 5, 6, 7]&[4]&True&[0.279, 0.217, 0.226, 0.151, 0.13]&[0.207, 0.116, 0.114, 0.042, 0.0]&e4\\\hline
\color{blue}Aspen-11&\color{blue}[23, 25, 26, 27]&\color{blue}[24]&\color{blue}True&\color{blue}[0.294, 0.264, 0.255, 0.23, 0.282]&\color{blue}[0.228, 0.187, 0.133, 0.081, 0.069]&\color{blue}e6\\\hline
Aspen-11&[33, 35, 36, 37]&[34]&False&[0.301, 0.248, 0.203, 0.204, 0.225]&[0.236, 0.172, 0.091, 0.052, 0.042]&f2\\\hline
Aspen-11&[43, 45, 46, 47]&[44]&False&[0.293, 0.24, 0.191, 0.16, 0.162]&[0.228, 0.162, 0.101, 0.061, 0.047]&f3\\\hline
Aspen-11&[24, 26, 27, 20]&[25]&False&[0.305, 0.286, 0.286, 0.288, 0.303]&[0.242, 0.207, 0.168, 0.131, 0.126]&f4\\\hline
Aspen-11&[34, 36, 37, 30]&[35]&False&[0.306, 0.266, 0.181, 0.167, 0.146]&[0.245, 0.18, 0.037, 0.035, 0.021]&f5\\\hline
Aspen-11&[4, 6, 7, 0]&[5]&False&[0.312, 0.243, 0.204, 0.174, 0.171]&[0.247, 0.171, 0.1, 0.037, 0.022]&f6\\\hline
\color{blue}Aspen-11&\color{blue}[23, 25, 26, 27]&\color{blue}[24]&\color{blue}True&\color{blue}[0.293, 0.262, 0.256, 0.239, 0.306]&\color{blue}[0.213, 0.171, 0.15, 0.092, 0.052]&\color{blue}f7\\\hline
Aspen-11&[33, 35, 36, 37]&[24]&True&[0.298, 0.24, 0.166, 0.146, 0.15]&[0.228, 0.153, 0.069, 0.026, 0.014]&f8\\\hline
Aspen-11&[43, 45, 46, 47]&[44]&True&[0.314, 0.281, 0.138, 0.122, 0.15]&[0.246, 0.201, 0.051, 0.02, 0.008]&f9\\\hline
Aspen-M-1&[34, 36, 37, 30]&[35]&True&[0.31, 0.279, 0.286, 0.254, 0.239]&[0.248, 0.206, 0.171, 0.081, 0.026]&g1\\\hline
Aspen-M-1&[30, 36, 35, 34]&[37]&True&[0.313, 0.281, 0.236, 0.213, 0.241]&[0.249, 0.204, 0.126, 0.043, 0.03]&g2\\\hline
Aspen-M-1&[35, 37, 30, 133]&[36]&True&[0.309, 0.294, 0.283, 0.223, 0.15]&[0.249, 0.218, 0.185, 0.073, 0.025]&g3\\\hline
Aspen-M-1&[15, 17, 114, 115]&[16]&True&[0.312, 0.289, 0.303, 0.231, 0.257]&[0.248, 0.217, 0.198, 0.093, 0.044]&g4\\\hline
Aspen-M-1&[115, 17, 16, 15]&[114]&True&[0.313, 0.292, 0.287, 0.271, 0.276]&[0.25, 0.221, 0.186, 0.112, 0.073]&g5\\\hline
Aspen-M-1&[23, 25, 26, 27]&[24]&True&[0.314, 0.269, 0.159, 0.137, 0.144]&[0.247, 0.197, 0.056, 0.015, 0.021]&g6\\\hline
Aspen-11&[23, 25, 26, 27]&[24]&False&[0.204, 0.19, 0.14, 0.162, 0.212]&[0.11, 0.086, 0.01, 0.015, 0.033]&h1\\\hline
Aspen-11&[23, 25, 26, 27]&[24]&False&[0.217, 0.184, 0.146, 0.184, 0.217]&[0.117, 0.09, 0.007, 0.017, 0.033]&h2\\\hline
\color{blue}Aspen-11&\color{blue}[23, 25, 26, 27]&\color{blue}[24]&\color{blue}True&\color{blue}[0.22, 0.152, 0.171, 0.179, 0.27]&\color{blue}[0.124, 0.055, 0.038, 0.003, 0.023]&\color{blue}h3\\\hline
Aspen-11&[23, 25, 26, 27]&[24]&True&[0.209, 0.161, 0.175, 0.163, 0.19]&[0.115, 0.048, 0.032, 0.0, 0.002]&h4\\\hline
\color{red}Aspen-11&\color{red}[5, 7, 0, 1]&\color{red}[6]&\color{red}True&\color{red}[0.223, 0.176, 0.162, 0.165, 0.241]&\color{red}[0.128, 0.08, 0.031, 0.0, 0.009]&\color{red}h5\\\hline
Aspen-11&[43, 45, 46, 47]&[44]&True&[0.204, 0.165, 0.151, 0.137, 0.138]&[0.112, 0.056, 0.032, 0.001, 0.002]&h6\\\hline
\color{red}Aspen-11&\color{red}[5, 7, 0, 1]&\color{red}[6]&\color{red}True&\color{red}[0.221, 0.168, 0.16, 0.161, 0.216]&\color{red}[0.129, 0.068, 0.031, 0.0, 0.007]&\color{red}h5B\\\hline
\color{red}Aspen-11&\color{red}[5, 7, 0, 1]&\color{red}[6]&\color{red}True&\color{red}[0.232, 0.175, 0.136, 0.172, 0.189]&\color{red}[0.134, 0.075, 0.008, 0.0, 0.001]&\color{red}h5C\\\hline
5q-qvm&[0, 2, 3, 4]&[1]&True&[0.316, 0.288, 0.281, 0.307, 0.349]&[0.25, 0.219, 0.177, 0.136, 0.086]&5qvm\\\hline

	\end{tabular}
\caption{Experimental results using different devices and qubits.  The data in blue (red) was used to generate Fig.~\ref{fig:experiment}(d) (Fig.~\ref{fig:exp2}). The last row shows the result of an ideal simulator with unitary gates, used to generate the theory curves.}
\label{table:exp}
\end{table}

\begin{figure}[h]
\includegraphics[scale=0.5]{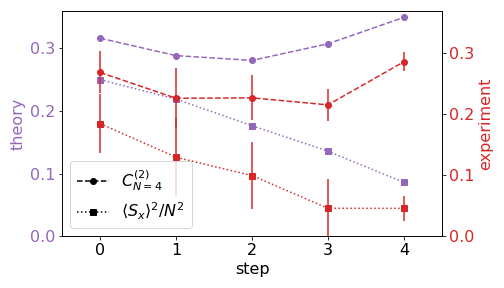}
\caption{Same as Fig.~\ref{fig:coherence}(c) using a different set of qubits, see data highlighted in blue in Table~\ref{table:exp}.}
\label{fig:exp2}
\end{figure}

\newpage

\section{Error mitigation}
\label{sec:mit}
\begin{figure}[h]
	\includegraphics[scale=0.5]{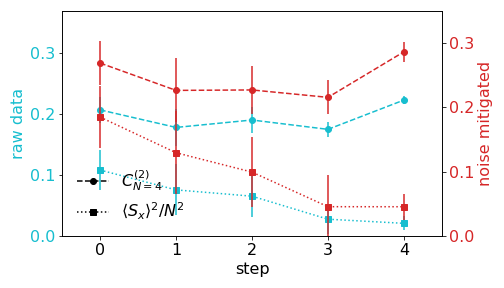}
	\caption{Same as Fig.~\ref{fig:experiment}(c), comparing the data with and without error mitigation.}
	\label{fig:mit}
\end{figure}

We use a simple version of error mitigation, aimed at reducing the effects of state-preparation and measurement (SPAM) errors. The protocol works in two steps: first, we calibrate the system qubit by preparing the 01010 and 10101 quantum states and measuring each qubit independently. From these measurements we compute the parameters $p_00$ and $p_11$ of each qubit, i.e. the probability that when the qubit was prepared in 0 it returns 0 and viceversa for 1. Next we perform our experiment as usual. Finally, we use the parameters $p_00$ and $p_11$ to correct the histograms of the experimental results. Technically, this is done by multiplying the histogram by the {\it inverse} of the confusion matrix $A$, defined as
\begin{align}
A = \left(\begin{matrix}p_{00} & 1-p_{11} \\1-p_{00} & p_{11}\end{matrix}\right)\times\left(\begin{matrix}p_{00} & 1-p_{11} \\1-p_{00} & p_{11}\end{matrix}\right)\times\left(\begin{matrix}p_{00} & 1-p_{11} \\1-p_{00} & p_{11}\end{matrix}\right)\times\left(\begin{matrix}p_{00} & 1-p_{11} \\1-p_{00} & p_{11}\end{matrix}\right)\times\left(\begin{matrix}p_{00} & 1-p_{11} \\1-p_{00} & p_{11}\end{matrix}\right),
\end{align}
where $\times$ denotes the tensor product. Fig.~\ref{fig:mit} demonstrates the functionality of this procedure.

\section{A minimal model of phase stiffness}
\label{sec:stiffness}
In this section we provide further details on the relation between number conservation and phase stiffness mentioned in the conclusion of the main text.

A system is said to have phase stiffness if its phase is stable to external perturbations. Following Landau, it is common to consider an impurity moving at a fixed velocity ${\bf v}$. Its effect can be studied by considering the co-moving frame, where $\omega_{\bf q} \to \omega'_{\bf q}=\omega_{\bf q} - {\bf v}\cdot {\bf q}$. Landau's criterion states that a system has phase stiffness if there exists a critical velocity $v^* > 0$ such that $\omega'_{\bf q}>0$ for all ${\bf q}$. A minimal model for phase stiffness is the $d$-dimensional field-theoretical Hamiltonian
\begin{align}
	H &= \int d^dx~ \frac{\rho_0}2\left[\partial_x\theta(x)\right]^2 + \frac{U}2 [n(x)-n_0]^2\label{eq:Hfield}
\end{align}
Here, $\phi(x)$ and $n(x)$ are, respectively the phase and number field operators, satisfying the canonical relation $[n(x'),\theta(x)]=i\delta(x-x')$. The first term of Eq.~(\ref{eq:Hfield}) describes the tendency of neighboring particles to acquire the same phase, for example, due to their kinetic energy. The second term of Eq.~(\ref{eq:Hfield}) attempts to reduce fluctuations of the number of partilces, for example, due to local interactions between the particles. The Hamiltonian (\ref{eq:Hfield}) can be easily diagonalized by moving to the momentum space, and its spectrum is $\omega_{\bf q} = \sqrt{\rho_0 U} q$. Hence, this system has phase stiffness with critical velocity $v^* = \sqrt{\rho_0 U}$. Importantly, $v^*$ tends to 0 for $U\to0$, showing that phase stiffness occurs only for interacting systems.

A simplified version of this criterion can be presented by considering a mean-field version of $H$, where the spatial derivative is substituted by the ratio between $\theta$ and the system size $L$. The resulting Hamiltonian is just an harmonic oscillator
\begin{align}
	H & = \frac{\rho_0L^{d-2}}{2}\theta^2 + \frac{U}{2L^d}(N-N_0)^2,\label{eq:simplified}
\end{align}
where $N=L^d n$ is the total number of particles and satisfies $[N,\theta]=i$. For $U=0$, the ground state of this Hamiltonian is the position state $|\theta = 0\rangle$. This state has a well defined global phase and corresponds to our $|\rm coherent\rangle$ state. The $U$ term leads to a ground state with a fluctuating $\theta$, along with suppressed fluctuations of $N$, analogous to $|\rm projected\rangle$. To evaluate the dynamical stability of a system, it is necessary to study its excitation spectrum: For $U=0$, the eigenstates of Eq.~(\ref{eq:simplified}) are given by $|\theta=\theta_0\rangle$, with eigenenergies $\rho_0^2L^{d-2}\theta_0^2/2$. Because these energies are arbitrarily small (for $\theta_0\to 0$), any weak perturbation can cause a shift in $\theta$, leading to a loss of phase coherence. In contrast, for $U\neq0$, the eigenenergies are quantized in units of $\sqrt{\rho_0 U}/L$, leading to a protection against external perturbations. Hence, the states  $|\rm projected\rangle$ and $|\rm coherent\rangle$ with a finite $L$, respectively, correspond to the ground states of systems with and without phase stiffness.

\end{document}